\documentclass[pdflatex,sn-basic,Numbered]{sn-jnl}

\usepackage{graphicx}
\usepackage{multirow}
\usepackage{amsmath,amssymb,amsfonts}
\usepackage{mathrsfs}
\usepackage[title]{appendix}
\usepackage{xcolor}
\usepackage{textcomp}
\usepackage{manyfoot}
\usepackage{booktabs}
\usepackage{algorithm}
\usepackage{algpseudocode}
\usepackage{listings}
\usepackage{array}
\usepackage{eso-pic}

\hypersetup{bookmarksdepth=2,pdftitle={Hyperscale Data Centers as Wide-Area Damping Actuators for Low-Inertia Power Systems},pdfauthor={Mohamed Shamseldein},pdfsubject={Author manuscript of published SGSE article; DOI 10.1007/s40866-026-00358-5}}

\begin{document}

\AddToShipoutPictureFG*{\AtPageUpperLeft{%
\put(\LenToUnit{35mm},\LenToUnit{-12mm}){%
\parbox[t]{140mm}{\normalfont\scriptsize
\textbf{Author manuscript.} Published as: M. Shamseldein,
\emph{Smart Grids and Sustainable Energy}, volume~11, article~43 (2026).
The Version of Record was published on 27~June~2026 and is available at
\url{https://doi.org/10.1007/s40866-026-00358-5}.
This author-prepared manuscript is not the publisher's Version of Record.
}}}}

\title{Hyperscale Data Centers as Wide-Area Damping Actuators for
Low-Inertia Power Systems}

\author*[1]{\fnm{Mohamed} \sur{Shamseldein}}\email{mohamed.shamseldein@eng.asu.edu.eg}

\affil*[1]{\orgdiv{Department of Electrical Power and Machines},
            \orgname{Faculty of Engineering, Ain Shams University},
            \orgaddress{\city{Cairo}, \country{Egypt}}}

\abstract{The decarbonization of electric power systems is displacing
synchronous generation with inverter-based renewable resources, eroding
the inertia and power-system-stabilizer-derived damping that historically
suppressed inter-area oscillations. In parallel, hyperscale
artificial-intelligence data centers have emerged as large, continuously
operated loads whose power draw is controllable on the millisecond
timescale through Dynamic Voltage and Frequency Scaling (DVFS). This
paper evaluates whether these compute loads can act as wide-area damping
actuators. The analysis uses a DAE-consistent bus-input matrix,
$\mathbf{B}_b=-\mathbf{T}^{-1}\mathbf{f}_y\mathbf{g}_y^{-1}\mathbf{g}_{u,b}$,
and validates the construction against the ANDES reduced state matrix and
a finite-difference input check. On the Kundur two-area benchmark the
DAE-consistent residue ranking preserves the main siting result: Bus~10 is
the highest-ranked non-generator bus for the 0.647~Hz inter-area mode.
A classical washout-plus-lead-lag controller raises damping from 3.43\%
to approximately 6\% using the same DAE input. Under this model, modern
10--50~ms battery and DVFS actuators require comparable
peak modulation for the same small-signal damping target. The data-center
wide-area controller still improves on a closest same-system
data-center-frequency-droop baseline: at comparable peak MW the
droop baseline reaches 4.76\% damping, and reaching 6\% requires
approximately twice the peak modulation. PMU-delay robustness is
also improved, with the damping crossing the open-loop value at
443~ms for the 10~ms DVFS actuator versus 374~ms for the 50~ms BESS
case in the Pad\'e-delay model. Saturation stress tests show that
ordinary 2~MW headroom does not bind for modal, fault-cleared-equivalent,
and line-outage-equivalent disturbances, while generator-trip-equivalent stress can bind
the actuator and must be treated as an operating constraint. IEEE 39-bus
scale-up confirms the DAE-based ranking procedure and highlights a practical
phase-feasibility issue: the best residue buses may require more phase
compensation than a single lead-lag stage can supply.}


\maketitle

\section{Introduction}

\subsection{Motivation: two decade-long trends converging}

The global electricity system is undergoing two simultaneous transformations
that together reshape the design space for frequency and oscillation
control. First, the displacement of synchronous thermal and hydro
generation by inverter-based renewable resources erodes the aggregate
system inertia and, with it, the damping traditionally provided by
synchronous machines and their power system stabilizers (PSSs)~\cite{nerc2023ibr,cigre2024inertia}.
In low-inertia systems, inter-area oscillations---the electromechanical
modes in which geographically separated generator groups swing against
each other at $0.1$--$1$~Hz---become harder to damp, and high renewable
penetration levels have been observed to excite modes that were
well-damped in pre-transition conditions. Wide-area damping control
(WADC) using phasor-measurement feedback has been developed over two
decades to address this challenge, typically via Flexible AC Transmission
System (FACTS) devices, high-voltage DC supplementary signals, or battery
energy storage systems~\cite{pnnl2018wadc,pal2005robustcontrol,kumar2025bess}.

Second, the rapid scaling of artificial-intelligence training and
inference workloads has created a new class of industrial loads:
hyperscale data centers with individual site power capacities of
$100$--$500$~MW and campus aggregations approaching $1$~GW. Global
data center electricity consumption is projected to exceed
$1{,}000$~TWh per year by 2026~\cite{iea2024ai}, and forecasts identify
approximately $1$~TW of installed hyperscale compute capacity by
$2030$~\cite{patel2024hyperscale}. These loads possess three
characteristics relevant for grid services: (i) their IT power is
modulable by up to $80\%$ within $10$~ms via Dynamic Voltage and
Frequency Scaling (DVFS) of the underlying processors; (ii) they operate
continuously, providing $24/7$ availability unlike discretionary
demand-response resources; and (iii) they are geographically distributed
by design, following fiber corridors and cheap-power regions, which
places many facilities at buses of non-trivial electrical relevance.

The convergence of these two trends raises a natural question: if the
low-inertia grid needs fast, distributed damping actuators, and the
hyperscale compute fleet already has them, can the compute fleet serve
as a wide-area damping resource? This paper develops a principled
methodology for that question and quantifies the performance of
DC-based wide-area damping against the nearest technological
alternative.

\paragraph{Sustainability framing.} The motivation above is grounded in
measured grid outcomes. In ERCOT, minimum synchronous inertia has
declined from $132$~GW$\cdot$s in 2013 to $115$~GW$\cdot$s in
2022~\cite{ercot2023inertia}, coinciding with more than $54$~GW of
installed inverter-based resources by the end of 2022. In Ireland,
non-synchronous penetration has exceeded $60\%$ on individual days and
TSOs curtailed approximately $1{,}795$~GWh of renewable output in 2023
to preserve system security~\cite{esig2024oscillations}. These are
concrete sustainability costs: each megawatt-hour of renewable
curtailment is a missed opportunity to displace fossil generation. To
the extent that data-center-based wide-area damping can relax stability
constraints that bind curtailment decisions, it contributes directly to
the decarbonization objective rather than to a narrow control-theoretic
metric. Recent independent assessments emphasize the same
opportunity: flexible AI data centers are increasingly characterized
as ``grid-interactive assets''~\cite{colangelo2026gridDC}, and integrating
even modest curtailability at scale would unlock tens of gigawatts of
new grid capacity without additional generation
capex~\cite{duke2024flexload}. The present paper contributes a specific
control-theoretic mechanism to that broader program.

\subsection{Related work}

\textbf{Wide-area damping control.} The problem of damping inter-area
oscillations via remote feedback has been studied since the early
1990s, with foundational methods based on residue analysis and
participation factors~\cite{larsen1981pss,larsen1981pss2,larsen1981pss3,wang1997residue,perez1982participation}.
Robust and adaptive designs followed, addressing time delay, mode
interaction, and parameter uncertainty~\cite{pal2005robustcontrol,chaudhuri2003widearea,sadabadi2021adaptive}.
Actuator choice has evolved in parallel: from PSSs on synchronous
generators~\cite{kundur1994stability}, to FACTS
devices~\cite{pal1999facts}, to HVDC supplementary
signals~\cite{rostamkolai1998hvdc}, to grid-forming inverters in
converter-dominated systems~\cite{oshea2024gfm,rokrok2024gfmpod,ippolitoMusca2024},
and most recently to BESS~\cite{kumar2025bess,singh2024bessdamping}.
Among these, BESS most directly competes with the load-based actuator
considered here.

\textbf{End-use loads for damping.} Liu~\textit{et~al.}~\cite{liu2014loaddamping}
first proposed that aggregated end-use loads could provide a WADC
function analogous to BESS. Subsequent work has extended the idea to
aggregated residential thermostatic loads~\cite{huang2016widearea},
industrial motor populations~\cite{gu2020industrial}, and electric
vehicle fleets~\cite{wang2023evdamping}. These studies share two
features: aggregation bandwidths are typically on the order of seconds
because of heterogeneous device dynamics, and the controlled loads are
indirectly actuated via price or frequency signals. Neither property
applies to a hyperscale data center, where the workload orchestrator is
a single authoritative controller with sub-second authority over CPU
and GPU power states across the entire campus.

\textbf{Data center demand response.} The potential for data centers to
participate in ancillary services has been recognized for over a
decade~\cite{ghatikar2012datacenter,wierman2014opportunities}. More
recent studies have demonstrated DC participation in fast frequency
response~\cite{alkez2021ffr,dvfs2006} and grid-level peak
shaving~\cite{liu2013datacenter}. None of these studies, to our
knowledge, has addressed inter-area oscillation damping---the problem
in which the distinct actuator properties of hyperscale DVFS
(millisecond response, geographic distribution, gigawatt-scale
capacity) are most differentiating.

\textbf{The gap.} Despite mature literatures on WADC, on load-based
damping, and on DC demand response, no prior work treats hyperscale
DCs as distributed wide-area damping controllers with a principled
siting-plus-control design methodology, and no prior work quantifies
the performance of DC-based damping against the operational
alternative of equivalent BESS. The present paper addresses both gaps.

\subsection{Contributions}

This paper makes four contributions, each validated on the canonical
Kundur 2-area benchmark with scale-up confirmation on IEEE 39-bus
New England.

\begin{enumerate}
\item A \emph{participation-factor-based siting methodology} that
ranks candidate DC locations by modal residue magnitude subject to a
phase-feasibility constraint. The methodology produces the
counter-intuitive result that area-proximate load buses outrank
tie-line buses for load-modulation actuators---a structural
departure from FACTS-based WADC siting wisdom, explained physically by
the differential area coupling of bus-level real-power injection
(Section~\ref{sec:methodology-siting-finding}, Table~\ref{tab:bus-ranking}).

\item A \emph{DAE-consistent bus-input model}. The inverse-reactance
expression is retained only as a screening proxy. All siting, controller,
and comparison results use the
ANDES Schur-complement input matrix and report the proxy error
(Section~\ref{sec:methodology-residue},
Table~\ref{tab:dae-ranking}).

\item A \emph{same-system comparison with both BESS and a prior-art
data-center droop analog}. The DAE-consistent simulations show that
10--50~ms BESS and DVFS actuators have comparable peak-MW requirements
for small-signal damping, but the proposed wide-area controller reaches
the 6\% target with about half the peak modulation of a local
frequency-droop data-center baseline
(Sections~\ref{sec:mw-comparison}--\ref{sec:prior-baseline}).

\item A \emph{constraint and scale-up assessment}. One-sided DVFS and
hard actuator limits are simulated directly in the reduced closed loop,
and IEEE~39-bus validation shows that phase feasibility can dominate
residue magnitude at larger scale (Sections~\ref{sec:saturation}
and~\ref{sec:ieee39-scaleup}).
\end{enumerate}

\subsection{Paper organization}

Section~\ref{sec:models} presents the power-system, data-center, BESS,
and wide-area-measurement models. Section~\ref{sec:methodology} develops
the participation-residue siting methodology and residue-method
controller design. Section~\ref{sec:controller} applies the methodology
to the Kundur 2-area benchmark and reports the nominal closed-loop
operating point. Section~\ref{sec:results} contains four validation
studies on Kundur followed by scale-up validation on IEEE~39-bus.
Section~\ref{sec:discussion} discusses scope, limitations, and
deployment. Section~\ref{sec:conclusion} concludes.

\section{System and Actuator Models}
\label{sec:models}

\subsection{Power system dynamic model}

We consider a multi-machine power system with $N_g$ synchronous
generators. The electromechanical dynamics of generator $i$ are
described by the classical swing equations:
\begin{align}
\frac{2H_i}{\omega_s} \frac{d\omega_i}{dt}
   &= P_{m,i} - P_{e,i} - D_i(\omega_i - \omega_s)\label{eq:swing1}\\
\frac{d\delta_i}{dt} &= \omega_i - \omega_s \label{eq:swing2}
\end{align}
where $H_i$ is the inertia constant on machine base, $\omega_i$ is the
rotor speed, $\delta_i$ is the rotor angle, $\omega_s = 2\pi f_s$ is the
synchronous angular frequency, $P_{m,i}$ is the mechanical power input,
$P_{e,i}$ is the electrical power output, and $D_i$ is the damping
coefficient. The electrical power $P_{e,i}$ couples all generators
through the network admittance matrix $\mathbf{Y}$ and the bus voltage
phasors.

For wide-area analysis, we define the Center-of-Inertia (COI) frequency:
\begin{equation}
f_{\mathrm{COI}} = \frac{\sum_{i=1}^{N_g} H_i f_i}{\sum_{i=1}^{N_g} H_i}
\label{eq:coi}
\end{equation}
and, for a two-area partition $\{\mathcal{A}, \mathcal{B}\}$, the
inter-area frequency signal:
\begin{equation}
\Delta\bar{\omega}_{\mathcal{AB}} =
\frac{1}{|\mathcal{A}|}\sum_{i \in \mathcal{A}} \omega_i
- \frac{1}{|\mathcal{B}|}\sum_{j \in \mathcal{B}} \omega_j
\label{eq:inter-area-signal}
\end{equation}
which serves as the remote measurement input to the wide-area damping
controller. All numerical experiments use the open-source ANDES
simulator~\cite{cui2021andes}, which solves the full differential-algebraic equation (DAE) system comprising (\ref{eq:swing1})--(\ref{eq:swing2})
together with exciter, turbine-governor, and network algebraic equations.

Linearization of the DAE system around an equilibrium
$(\mathbf{x}_0, \mathbf{y}_0)$ yields the reduced state-space form:
\begin{align}
\dot{\mathbf{x}} &= \mathbf{A}\,\mathbf{x} + \mathbf{B}\,u \label{eq:ss-dyn}\\
y &= \mathbf{C}\,\mathbf{x} \label{eq:ss-out}
\end{align}
where
$\mathbf{A} = \mathbf{f}_x - \mathbf{f}_y\mathbf{g}_y^{-1}\mathbf{g}_x$
is the reduced-order state matrix, $\mathbf{x}$ collects the differential
states (rotor angles, rotor speeds, exciter and governor states), $u$ is
the scalar control input (active-power modulation at the data center
bus), and $y$ is the remote inter-area speed signal
(\ref{eq:inter-area-signal}). The input matrix $\mathbf{B}$ encodes the
network-propagation of a bus-level power injection into the swing
dynamics of each generator; its construction for a specific DC siting
bus $b$ is given in Section~\ref{sec:methodology}.

\subsection{Data center DVFS actuator model}

Hyperscale data centers modulate power consumption through Dynamic
Voltage and Frequency Scaling (DVFS), which adjusts CPU clock frequency
and supply voltage in concert. For a CMOS processor the instantaneous
IT power follows a power-law relationship:
\begin{equation}
P_{\mathrm{IT}}(\alpha) = P_{\mathrm{base}}\,\alpha^k
\label{eq:dvfs-power}
\end{equation}
where $\alpha \in [\alpha_{\min}, 1]$ is the throttle ratio,
$P_{\mathrm{base}}$ is the rated IT load, and $k \approx 2.5$ captures
the combined voltage-frequency scaling of modern server-class
processors~\cite{barroso2018datacenter}. A minimum throttle
$\alpha_{\min} = 0.5$ is imposed by architectural guard-bands on voltage
margin and thermal slew. Total facility power is approximated as
$P_{\mathrm{DC}} = \mathrm{PUE}\cdot P_{\mathrm{IT}}$, with PUE in the
range $1.15$--$1.4$ for modern hyperscale facilities.

The maximum downward active-power modulation is:
\begin{equation}
\Delta P_{\max} = \mathrm{PUE}\cdot P_{\mathrm{base}}\,(1 - \alpha_{\min}^k)
\approx 0.82\,\mathrm{PUE}\,P_{\mathrm{base}}
\label{eq:max-flex}
\end{equation}
For a 200~MW hyperscale campus with $\mathrm{PUE} = 1.2$, this yields
approximately $197$~MW of fast-acting flexibility.

The DVFS actuator itself responds with first-order dynamics:
\begin{equation}
\tau_{\mathrm{DVFS}}\,\frac{d\alpha}{dt} = \alpha_{\mathrm{cmd}} - \alpha
\label{eq:dvfs-dyn}
\end{equation}
where $\tau_{\mathrm{DVFS}} = 10$~ms is used as the nominal value. This
represents an optimistic but physically plausible aggregate of the
power-management controller, voltage-regulator response, and
scheduler acknowledgment~\cite{bansal2013dvfs}. The comparison
therefore tests actuator-time-constant sensitivity rather than relying
on this value as a fixed advantage.

For small-signal analysis around a nominal throttle $\alpha_0$,
linearization of (\ref{eq:dvfs-power})--(\ref{eq:dvfs-dyn}) yields:
\begin{equation}
\Delta P_{\mathrm{IT}} \approx
P_{\mathrm{base}}\,k\,\alpha_0^{k-1}\,\Delta\alpha
\label{eq:dvfs-linear}
\end{equation}
with actuator response
$\frac{\Delta\alpha(s)}{\Delta\alpha_{\mathrm{cmd}}(s)} =
\frac{1}{1 + s\tau_{\mathrm{DVFS}}}$.

\subsection{BESS actuator model (comparison baseline)}

For a like-for-like comparison we model a BESS with identical power
rating and control architecture. The electrical response is first-order:
\begin{equation}
\tau_{\mathrm{BESS}}\,\frac{dP_{\mathrm{BESS}}}{dt}
   = P_{\mathrm{cmd}} - P_{\mathrm{BESS}}
\label{eq:bess-dyn}
\end{equation}
with $\tau_{\mathrm{BESS}} = 50$~ms used as the nominal baseline and
$\tau_{\mathrm{BESS}}\in\{10,20,50,100\}$~ms tested in sensitivity
analysis. The 50~ms case represents a conservative inverter-plus-plant
response, while 10--20~ms represents modern fast-converter capability.
Unlike DVFS, BESS modulation is symmetric around zero power; the
state-of-charge constraint is not binding for the short-duration
($<25$~s) oscillation damping task studied here and is omitted from the
small-signal analysis.

\subsection{Wide-area measurement model}

The damping controller uses the inter-area frequency signal
$\Delta\bar{\omega}_{\mathcal{AB}}$ obtained from phasor measurement
units (PMUs) distributed across both areas. Wide-area communication
introduces an aggregate transport delay $\tau_{\mathrm{PMU}}$, typically
$20$--$100$~ms for a regional synchrophasor network~\cite{chakrabortty2011wams}.
For state-space analysis we approximate the pure delay with a first-order
Pad\'e expansion:
\begin{equation}
e^{-s\tau_{\mathrm{PMU}}}
  \approx \frac{1 - s\tau_{\mathrm{PMU}}/2}{1 + s\tau_{\mathrm{PMU}}/2}
\label{eq:pade}
\end{equation}
which adds a single state to the closed-loop system and preserves the
phase-delay behavior up to $\omega \tau_{\mathrm{PMU}} \sim 1$.

\subsection{Summary of model parameters}

Table~\ref{tab:params} collects the model parameters used throughout.

\begin{table}[t]
\caption{Model parameters.}
\label{tab:params}
\centering
\begin{tabular}{lcl}
\toprule
Parameter & Value & Notes \\
\midrule
DVFS exponent $k$                        & $2.5$   & CMOS scaling \\
Minimum throttle $\alpha_{\min}$         & $0.5$   & architectural guard \\
DVFS time constant $\tau_{\mathrm{DVFS}}$ & $10$~ms & nominal (optimistic) \\
Power Usage Effectiveness PUE            & $1.2$   & typical facility \\
BESS actuator lag $\tau_{\mathrm{BESS}}$ & $10$--$100$~ms & sensitivity range \\
PMU latency $\tau_{\mathrm{PMU}}$         & $20$--$100$~ms & WAMS range \\
Washout time constant $T_w$              & $10$~s  & controller design \\
Synchronous frequency $f_s$              & $60$~Hz & \\
\bottomrule
\end{tabular}
\end{table}

\section{Methodology: Participation-Residue Analysis for DC Siting and
Controller Design}
\label{sec:methodology}

The methodology proceeds in four stages: (i) small-signal mode
identification, (ii) area decomposition from mode shape, (iii)
residue-based ranking of candidate DC-siting buses, and (iv) design of
a phase-compensated damping controller. The full procedure is summarized
in Algorithm~\ref{alg:siting} (siting) and Algorithm~\ref{alg:controller}
(controller design).

\subsection{Small-signal mode identification}

From the reduced state matrix $\mathbf{A} \in \mathbb{R}^{n \times n}$
defined in (\ref{eq:ss-dyn}), the right and left eigendecompositions are:
\begin{equation}
\mathbf{A}\mathbf{v}_k = \lambda_k\,\mathbf{v}_k,
\qquad
\mathbf{w}_k^\top \mathbf{A} = \lambda_k\,\mathbf{w}_k^\top
\label{eq:eigendecomp}
\end{equation}
with bi-orthogonality $\mathbf{w}_k^\top \mathbf{v}_k = 1$. Each
complex eigenvalue $\lambda_k = \sigma_k + j\omega_k$ defines a mode
with natural frequency $f_k = |\omega_k|/(2\pi)$ and damping ratio:
\begin{equation}
\zeta_k = -\sigma_k\, \big/\, \sqrt{\sigma_k^2 + \omega_k^2}
\label{eq:damping}
\end{equation}
The \emph{participation factor} of state $i$ in mode $k$ is:
\begin{equation}
p_{ki} = v_{ki}\,w_{ki}
\label{eq:pfactor}
\end{equation}
and satisfies $\sum_i p_{ki} = 1$. We classify a mode $k$ as
\emph{electromechanical} if its rotor-state participation
$\sum_{i \in \mathcal{R}} |p_{ki}| > 0.3$, where $\mathcal{R}$ indexes
the $\{\delta_i, \omega_i\}$ states of all synchronous machines.

\subsection{Area decomposition from mode shape}

For an inter-area mode, the omega eigenvector $v_{ki}^\omega$ carries
phase information that partitions the generator set into coherent groups.
We define the phase of generator $i$ in mode $k$ relative to a reference
generator $i^\star$:
\begin{equation}
\phi_{i}^{(k)} = \arg\big(v_{ki}^\omega / v_{k i^\star}^\omega\big)
\label{eq:phase}
\end{equation}
and assign $i$ to area $\mathcal{A}$ if $|\phi_{i}^{(k)}| < \pi/2$, else
to area $\mathcal{B}$. For the Kundur 2-area benchmark at $0.647$~Hz,
this yields $\mathcal{A} = \{G_1, G_2\}$ and $\mathcal{B} = \{G_3, G_4\}$.
The inter-area output matrix is:
\begin{equation}
\mathbf{C} = \frac{1}{|\mathcal{A}|}\sum_{i \in \mathcal{A}}
                \mathbf{e}_{\omega_i}^\top
            - \frac{1}{|\mathcal{B}|}\sum_{j \in \mathcal{B}}
                \mathbf{e}_{\omega_j}^\top
\label{eq:cmatrix}
\end{equation}
where $\mathbf{e}_{\omega_i}$ selects the $\omega_i$ state.

\subsection{Residue-based bus ranking}
\label{sec:methodology-residue}

For a candidate DC siting at bus $b$, the input matrix is constructed
directly from the DAE linearization. ANDES represents
the initialized system as
\begin{align}
\mathbf{T}\dot{\mathbf{x}} &= \mathbf{f}(\mathbf{x},\mathbf{y}) \\
\mathbf{0} &= \mathbf{g}(\mathbf{x},\mathbf{y}) + \mathbf{g}_{u,b}u_b
\end{align}
where $u_b$ is incremental active-power load at bus $b$ and
$\mathbf{g}_{u,b}$ selects the active-power balance row of that bus.
Thus $u_b>0$ denotes increased data-center consumption, equivalently
negative net active-power injection into the network. All residues,
controller gains, and actuator limits use this load-positive convention.
Eliminating algebraic perturbations gives
\begin{equation}
\mathbf{B}_b
= -\mathbf{T}^{-1}\mathbf{f}_y\mathbf{g}_y^{-1}\mathbf{g}_{u,b}.
\label{eq:dae-input}
\end{equation}
The reduced state matrix is computed consistently as
\begin{equation}
\mathbf{A}
=\mathbf{T}^{-1}\left(\mathbf{f}_x-\mathbf{f}_y\mathbf{g}_y^{-1}\mathbf{g}_x\right).
\label{eq:dae-a}
\end{equation}
In the Kundur case, rebuilding (\ref{eq:dae-a}) from the ANDES Jacobian
matches \texttt{ss.EIG.As} with maximum absolute error below
$10^{-13}$, and a central finite-difference check of
(\ref{eq:dae-input}) at Bus~10 gives relative error below $10^{-14}$.

For comparison only, we also retain the inverse-reactance
screen. Denoting $\mathbf{X}=\Im(\mathbf{Y}^{-1})$, that proxy assigns
generator fractions
\begin{equation}
\alpha_{ib} = \frac{1/|X_{b,b_i}|}{\sum_{j} 1/|X_{b,b_j}|}.
\label{eq:alpha}
\end{equation}
This proxy is used only for screening, not for the final controller design.

The \emph{modal residue} of mode $k$ for input at bus $b$ is:
\begin{equation}
R_{k,b} = (\mathbf{C}\,\mathbf{v}_k)\,(\mathbf{w}_k^\top\,\mathbf{B}_b)
\label{eq:residue}
\end{equation}
$R_{k,b}$ captures both the achievable damping effect (its magnitude
$|R_{k,b}|$) and the phase compensation required of any feedback
controller (its argument $\arg R_{k,b}$). Buses are ranked by
$|R_{k,b}|$ subject to the phase-feasibility constraint
$|\arg R_{k,b}| < \phi_{\max}$ imposed by the controller structure.
Generator terminal buses are excluded.

\subsection{A counter-intuitive siting finding}
\label{sec:methodology-siting-finding}

Table~\ref{tab:bus-ranking} presents the DAE-consistent ranking for the
Kundur $0.647$~Hz mode. The generator terminal buses are
excluded by the siting rule. Among non-generator buses, Bus~10 remains
highest, followed by Bus~5. Tie-line buses $7$, $8$, and $9$ rank lower.
This structural siting result is therefore tied to the DAE input matrix
rather than the inverse-reactance proxy.

\begin{table}[t]
\caption{DAE-consistent bus residue ranking for DC siting on the Kundur
$0.647$~Hz inter-area mode. Generator terminal buses occupy the omitted ranks and are excluded by the siting rule; the six non-generator buses are shown. Positive input denotes incremental load.}
\label{tab:bus-ranking}
\centering
\begin{tabular}{cccl}
\toprule
Rank & Bus & $|R|$ & $\arg R$ (deg) \\
\midrule
3    & 10  & $1.94\times10^{-3}$ & $-0.50$ \quad (Area B side) \\
4    & 5   & $1.82\times10^{-3}$ & $+179.70$ \quad (Area A side) \\
7    & 9   & $1.30\times10^{-3}$ & $+0.76$ \quad (Tie-line side) \\
8    & 6   & $1.19\times10^{-3}$ & $-179.05$ \quad (Tie-line side) \\
9    & 8   & $7.90\times10^{-4}$ & $+2.53$ \quad (Tie-line) \\
10   & 7   & $6.99\times10^{-4}$ & $-177.36$ \quad (Tie-line) \\
\bottomrule
\end{tabular}
\end{table}

\begin{table}[t]
\caption{DAE input validation and proxy comparison on Kundur. The proxy
correlation is across all ten buses.}
\label{tab:dae-ranking}
\centering
\begin{tabular}{lc}
\toprule
Quantity & Value \\
\midrule
$\max|\mathbf{A}_{\mathrm{rebuilt}}-\mathbf{A}_{\mathrm{ANDES}}|$ & $9.8\times10^{-14}$ \\
Bus~10 finite-difference input relative error & $1.2\times10^{-15}$ \\
Best non-generator bus under DAE input & Bus~10 \\
Spearman correlation of proxy and DAE $|R|$ & $0.636$ \\
Proxy role in final design & Screening only \\
\bottomrule
\end{tabular}
\end{table}

The explanation is physical: a DC injection at a tie-line bus has
approximately equal electrical coupling to generators in both areas.
Because the two areas swing in opposite phase during an inter-area
oscillation, the resulting damping torques on $\mathcal{A}$ and
$\mathcal{B}$ partially cancel in (\ref{eq:residue}). At a
load-proximate bus like Bus~10 the injection couples preferentially to
its own area, producing a coherent damping torque. This diverges from
classical FACTS/HVDC wisdom because those devices modulate \emph{tie-line
flow}, whereas DC load modulation injects \emph{bus-level real power}.
Bus~10 is selected as the reference DC siting location throughout
Section~\ref{sec:results}.

\subsection{Residue-method damping controller}

The controller is a classical three-stage structure: washout, phase-compensating lead-lag, and gain:
\begin{equation}
H(s) = K \cdot \frac{s T_w}{1 + s T_w}
          \cdot \frac{1 + s T_1}{1 + s T_2}
\label{eq:controller}
\end{equation}
The washout ($T_w = 10$~s) blocks the DC component. The lead-lag
provides phase shift to align the controller output with the residue
direction. For closed-loop feedback $u = -H(s)\,y$, the perturbation
of the target eigenvalue under small gain is:
\begin{equation}
\Delta\lambda_k \approx -K\,H_0(j\omega_k)\,R_{k,b}
\label{eq:eig-perturb}
\end{equation}
where $H_0(s)$ is the unity-gain part of (\ref{eq:controller}). For
maximum leftward migration (pure damping improvement with minimal
frequency shift), the product $H_0(j\omega_k)\,R_{k,b}$ must be real
and positive:
\begin{equation}
\arg H_0(j\omega_k) = -\arg R_{k,b}
\label{eq:phase-req}
\end{equation}
Setting $a = T_1/T_2$ and centering the maximum-phase-shift frequency
at $\omega_k$:
\begin{equation}
a = \frac{1 + \sin|\phi_{\mathrm{LL}}|}{1 - \sin|\phi_{\mathrm{LL}}|},
\quad T_2 = \frac{1}{\omega_k \sqrt{a}},
\quad T_1 = a T_2
\label{eq:leadlag}
\end{equation}
with $T_1, T_2$ swapped if $\phi_{\mathrm{LL}} < 0$. A single-stage
lead-lag provides phase shift up to $\phi_{\max} \approx 65^\circ$ at
the practical ratio $a_{\max} = 20$, defining the phase-feasibility
constraint.

Finally, $K$ is selected to achieve target closed-loop damping ratio
$\zeta^{\star}$ by sweeping and tracking the mode:
\begin{equation}
K^{\star} = \arg\min_{K}\ \big|\zeta_k^{\mathrm{CL}}(K) - \zeta^{\star}\big|
\label{eq:gain-select}
\end{equation}
subject to numerical stability.

\begin{algorithm}[t]
\caption{DC Siting by Residue Ranking}
\label{alg:siting}
\begin{algorithmic}[1]
\Require Linearized system $\mathbf{A}$, buses, generators, target mode index $k^\star$
\Ensure Best siting bus $b^\star$, input matrix $\mathbf{B}_{b^\star}$
\State Compute eigendecomposition of $\mathbf{A}$; extract $\lambda_{k^\star}, \mathbf{v}_{k^\star}, \mathbf{w}_{k^\star}$
\State Classify generators into areas $\mathcal{A}, \mathcal{B}$ via (\ref{eq:phase}); build $\mathbf{C}$ via (\ref{eq:cmatrix})
\ForAll{non-generator buses $b$}
    \State Construct $\mathbf{B}_b$ from the DAE reduction (\ref{eq:dae-input})
    \State Compute residue $R_{k^\star,b}$ via (\ref{eq:residue})
\EndFor
\State $\mathcal{F} \gets \{b : |\arg R_{k^\star,b}| < \phi_{\max}\}$
\State \textbf{return} $b^\star = \arg\max_{b \in \mathcal{F}} |R_{k^\star,b}|$
\end{algorithmic}
\end{algorithm}

\begin{algorithm}[t]
\caption{Damping Controller Design}
\label{alg:controller}
\begin{algorithmic}[1]
\Require Target mode $\lambda_k$, residue $R_{k^\star,b^\star}$, actuator time constant $\tau_{\mathrm{act}}$, target damping $\zeta^\star$
\Ensure Controller parameters $(K, T_w, T_1, T_2)$
\State Set $T_w = 10$~s
\State Compute $\phi_{\mathrm{LL}}$ via (\ref{eq:phase-req}); obtain $T_1, T_2$ via (\ref{eq:leadlag})
\State Sweep $K$ over $[10, 10^5]$; for each $K$ build closed-loop $\mathbf{A}_{\mathrm{cl}}$ and find eigenvalue closest to $\lambda_k$
\State Record $(K_j, \zeta_j^{\mathrm{CL}})$ pairs
\State $K^\star \gets \arg\min_j |\zeta_j^{\mathrm{CL}} - \zeta^\star|$
\State \textbf{return} $(K^\star, T_w, T_1, T_2)$
\end{algorithmic}
\end{algorithm}

\section{Controller Design and Closed-Loop Validation on Kundur}
\label{sec:controller}

\subsection{Testbench}

The Kundur two-area system~\cite{kundur1994stability} is chosen for
(i) a single well-documented inter-area mode near $0.6$~Hz with published
$2$--$5\%$ damping; (ii) two local modes near $1.1$~Hz enabling clean
multi-mode analysis; (iii) complete data shipped with
ANDES~\cite{cui2021andes} (\texttt{kundur\_full.xlsx}: four 900~MVA
GENROU machines with EXDC2 exciters and TGOV1 governors); and
(iv) a compact 10-bus topology making bus ranking inspectable by hand.

Power flow and time-domain initialization yield the open-loop
$\mathbf{A} \in \mathbb{R}^{52 \times 52}$. Three electromechanical modes
emerge; see Table~\ref{tab:kundur-modes}.

\begin{table}[t]
\caption{Kundur 2-area open-loop electromechanical modes.}
\label{tab:kundur-modes}
\centering
\begin{tabular}{lccl}
\toprule
Mode & $f$ (Hz) & $\zeta$ (\%) & Type \\
\midrule
Inter-area & $0.647$ & $3.43$ & $\{G_1,G_2\}$ vs $\{G_3,G_4\}$ \\
Local A1   & $1.108$ & $8.66$ & $G_1$ vs $G_2$ \\
Local A2   & $1.141$ & $8.86$ & $G_3$ vs $G_4$ \\
\bottomrule
\end{tabular}
\end{table}

\subsection{Bus selection and controller parameters}

Applying Algorithm~\ref{alg:siting}, Bus~10 ranks highest among non-generator buses ($|R|=1.94\times10^{-3}$,
$\arg R = -0.50^\circ$ under the positive-load sign convention).
Applying Algorithm~\ref{alg:controller} with target
$\zeta^\star = 6\%$:
\begin{equation}
T_w = 10~\mathrm{s},\;\;
T_1 = 0.242~\mathrm{s},\;\;
T_2 = 0.250~\mathrm{s},\;\;
K^\star = 5.36 \times 10^{1}
\label{eq:kundur-params}
\end{equation}
For the BESS baseline with $\tau_{\mathrm{BESS}}=50$~ms, the same
structural parameters are used but with $K^\star_{\mathrm{BESS}} = 5.77
\times 10^{1}$ for comparable damping.

\subsection{Closed-loop eigenvalue behavior}

\begin{figure}[t]
\centering
\includegraphics[width=\linewidth]{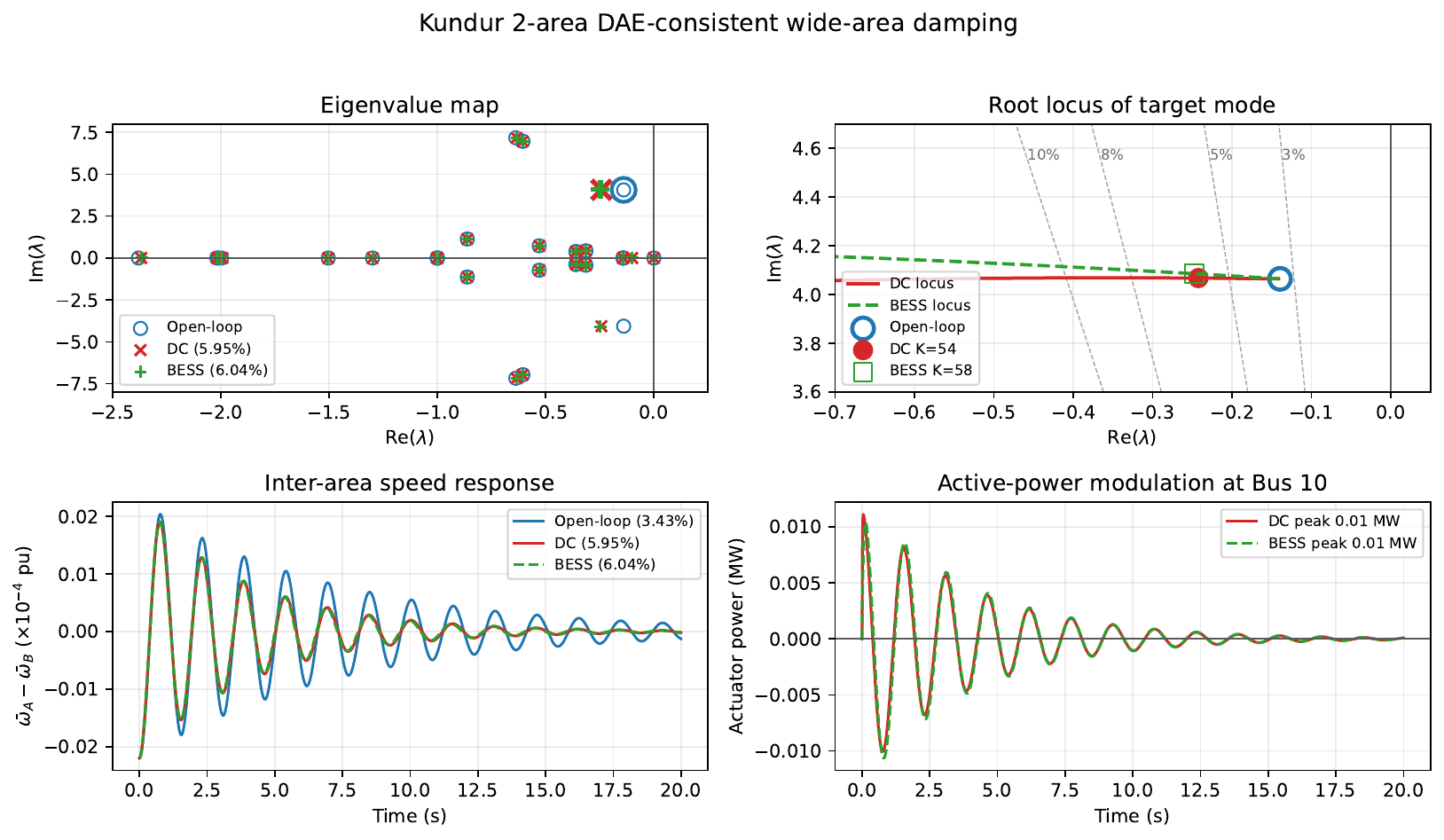}
\caption{Kundur 2-area: DC vs BESS wide-area damping at Bus~10.
(a) Eigenvalue map with open-loop and closed-loop poles overlaid.
(b) Root locus of the inter-area mode as $K$ increases.
(c) Time-domain inter-area speed signal under unit-eigenvector
excitation (open-loop/DC/BESS).
(d) Active-power modulation profile at Bus~10.}
\label{fig:core-damping}
\end{figure}

Figure~\ref{fig:core-damping} shows the validation. The inter-area mode
migrates from $\lambda_{\mathrm{OL}} = -0.140 + j4.065$ to
$\lambda_{\mathrm{CL}} \approx -0.242 + j4.067$:
\begin{equation}
f:\;0.647\,\mathrm{Hz} \to 0.647\,\mathrm{Hz},\quad
\zeta:\;3.43\% \to 5.95\%
\label{eq:damping-shift}
\end{equation}
The frequency shift is below $0.1\%$, indicating operation near the
damping-only regime. For the normalized 1\% modal initial condition used
in the simulation protocol, the peak DC active-power modulation is
$0.011$~MW on a $100$~MVA system base; the 50~ms BESS case is comparable
at $0.011$~MW. Thus the DAE-consistent model indicates comparable BESS
and DVFS peak-MW requirements for this small-signal comparison.

\section{Experimental Results}
\label{sec:results}

\subsection{Actuator-lag sensitivity: peak power comparison}
\label{sec:mw-comparison}

\begin{figure}[t]
\centering
\includegraphics[width=\linewidth]{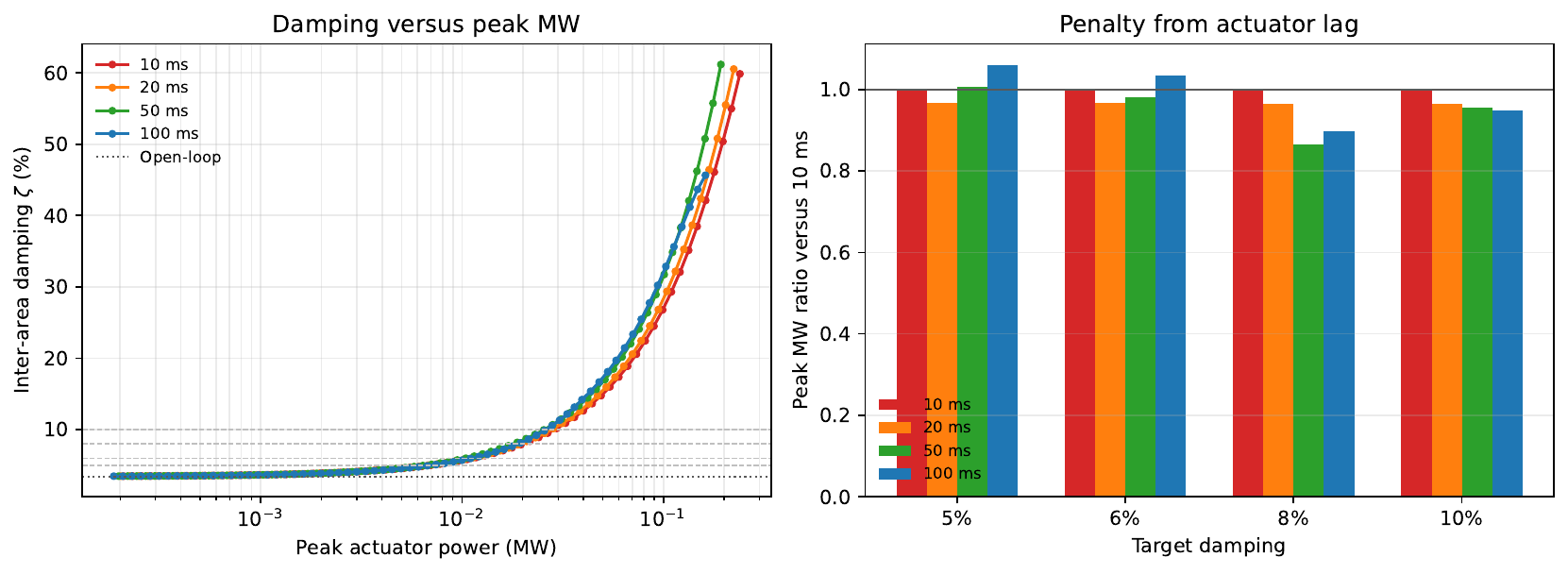}
\caption{Actuator-time-constant sensitivity on Kundur. (a) Damping
ratio vs peak actuator power for 10--100~ms actuator lags. (b) Peak-MW
ratio relative to the 10~ms case.}
\label{fig:mw-sensitivity}
\end{figure}

For each target closed-loop damping
$\zeta^\star \in \{5\%, 6\%, 8\%, 10\%\}$, we tune $K$ via
Algorithm~\ref{alg:controller} and simulate the $20$~s response to the
same 1\% inter-area-mode initial condition. Peak actuator power is
recorded. Figure~\ref{fig:mw-sensitivity} presents the
damping-versus-MW curves for 10, 20, 50, and 100~ms actuator lags;
Table~\ref{tab:mw-ratio} summarizes the target points.

\begin{table}[t]
\caption{Peak actuator power required for target damping with the
DAE-consistent input matrix (Kundur, 1\% modal initial condition).
The 10~ms column represents the nominal fast-DVFS case; 50~ms is the
nominal BESS baseline.}
\label{tab:mw-ratio}
\centering
\begin{tabular}{ccccc}
\toprule
$\zeta^\star$ & 10~ms & 20~ms & 50~ms & 100~ms \\
\midrule
$5\%$  & $0.0070$ & $0.0068$ & $0.0070$ & $0.0074$ \\
$6\%$  & $0.0117$ & $0.0113$ & $0.0115$ & $0.0121$ \\
$8\%$  & $0.0217$ & $0.0209$ & $0.0188$ & $0.0195$ \\
$10\%$ & $0.0295$ & $0.0284$ & $0.0281$ & $0.0280$ \\
\bottomrule
\end{tabular}
\end{table}

For the small-signal modal excitation, actuator lags from 10 to 100~ms all reach
the same damping targets with comparable peak modulation. This is
expected because the DAE-consistent Bus~10 residue has near-zero phase;
the first-order actuator lag at 0.647~Hz is not large enough by itself
to create a large BESS peak-MW penalty.

\subsection{Robustness to PMU communication latency}
\label{sec:pmu-latency}

\begin{figure}[t]
\centering
\includegraphics[width=\linewidth]{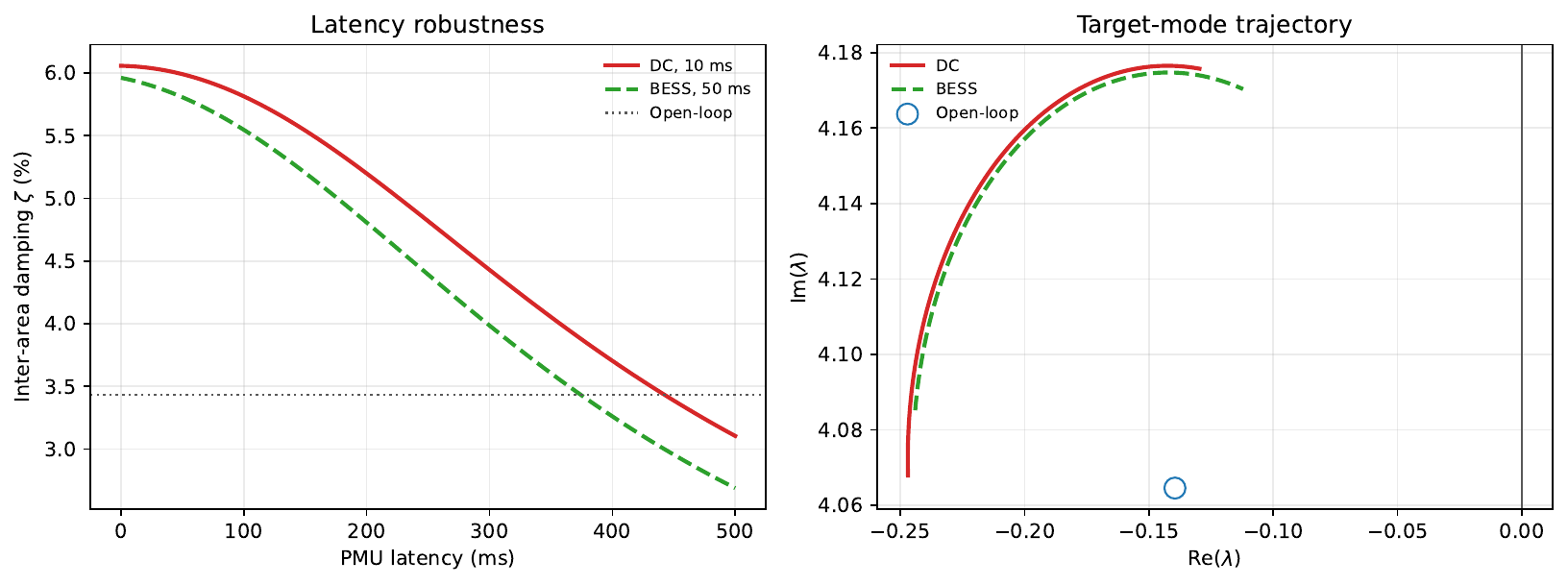}
\caption{PMU latency robustness on Kundur. (a) Damping vs latency
with open-loop baseline and instability region.
(b) Inter-area mode trajectory as $\tau_{\mathrm{PMU}}$ increases.}
\label{fig:pmu-latency}
\end{figure}

We insert (\ref{eq:pade}) into the measurement path and sweep
$\tau_{\mathrm{PMU}}$ from $0$ to $500$~ms holding other parameters
fixed. Figure~\ref{fig:pmu-latency} plots the results; Table~\ref{tab:latency}
tabulates key points.

\begin{table}[t]
\caption{Damping ratio vs PMU latency.}
\label{tab:latency}
\centering
\begin{tabular}{ccc}
\toprule
$\tau_{\mathrm{PMU}}$ (ms) & DC $\zeta$ (\%) & BESS $\zeta$ (\%) \\
\midrule
$0$   & $6.06$ & $5.96$ \\
$50$  & $5.99$ & $5.81$ \\
$100$ & $5.81$ & $5.55$ \\
$300$ & $4.43$ & $3.98$ \\
$500$ & $3.10$ & $2.69$ \\
\bottomrule
\end{tabular}
\end{table}

Damping drops to the open-loop baseline ($3.43\%$) at
$\tau_{\mathrm{PMU}} = 443$~ms for the 10~ms DVFS actuator versus
$374$~ms for the 50~ms BESS case, a $1.18\times$ latency margin in this
Pad\'e-delay model. The improvement is real but substantially smaller
than a simple actuator-lag-only estimate would suggest.

\subsection{Same-system prior-art data-center baseline}
\label{sec:prior-baseline}

Most prior data-center grid-service work addresses fast frequency
response or load following, not inter-area damping. We therefore add the
closest same-system analog: a local frequency-droop controller at the
same Bus~10 actuator, through the same DAE input matrix and 10~ms
actuator lag, but without the remote inter-area signal or residue-based
phase design.
Specifically,
\begin{equation}
\tau_a \dot u_d = K_d\,\Delta\bar{\omega}_B - u_d,\qquad
\Delta\bar{\omega}_B =
\frac{1}{|\mathcal{G}_B|}\sum_{i\in\mathcal{G}_B}\Delta\omega_i,
\label{eq:local-droop}
\end{equation}
with $\tau_a=10$~ms and $u_d>0$ increasing the Bus~10 load. The droop
gain $K_d$ is swept over $10^0$--$10^3$; Table~\ref{tab:prior-baseline}
reports the closest point to the proposed WADC peak MW and the
minimum-MW point reaching approximately 6\% damping.

\begin{table}[t]
\caption{Same-system comparison against a local data-center droop
baseline on Kundur.}
\label{tab:prior-baseline}
\centering
\begin{tabular}{lccc}
\toprule
Case & $\zeta$ (\%) & Peak MW & Notes \\
\midrule
Open loop & $3.43$ & -- & -- \\
Local droop, comparable peak MW & $4.76$ & $0.0116$ & local frequency only \\
Local droop tuned to $\sim6\%$ & $6.12$ & $0.0222$ & $1.97\times$ WADC MW \\
Proposed WADC & $5.99$ & $0.0113$ & remote inter-area signal \\
\bottomrule
\end{tabular}
\end{table}

\begin{figure}[t]
\centering
\includegraphics[width=\linewidth]{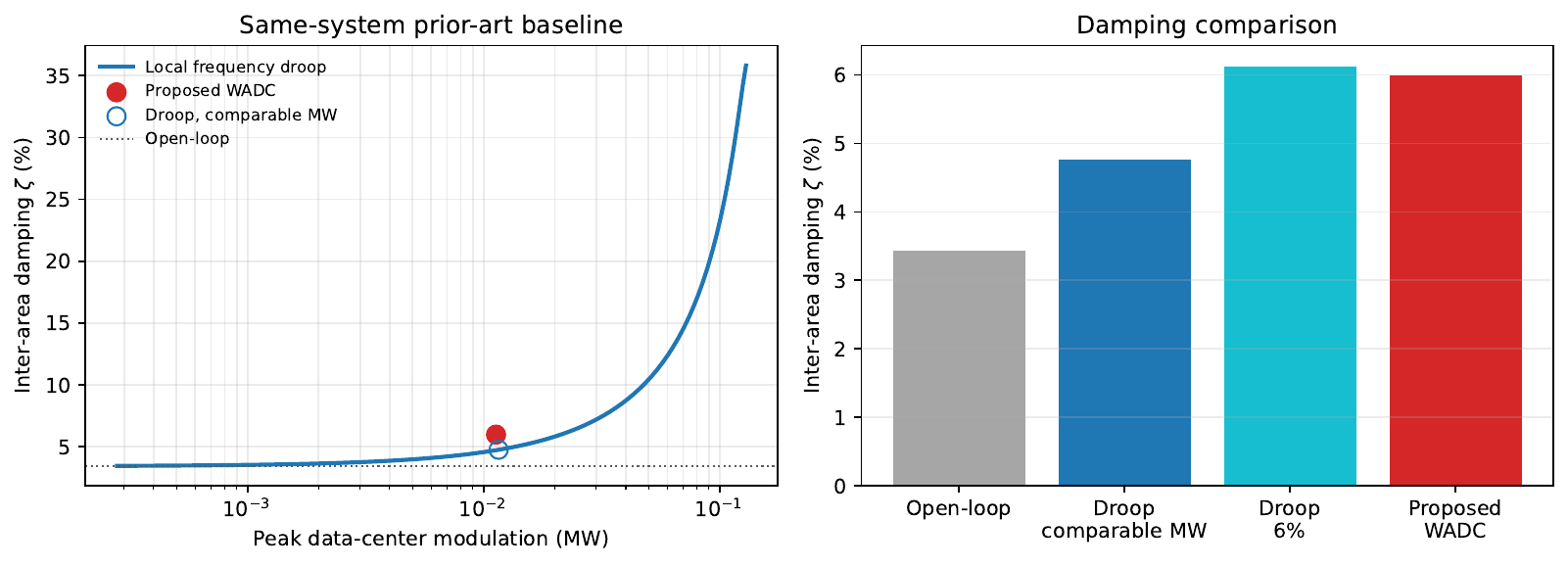}
\caption{Closest same-system prior-art baseline. The local
frequency-droop data-center controller improves damping, but reaching
the same 6\% target requires approximately twice the peak modulation of
the proposed wide-area controller.}
\label{fig:prior-baseline}
\end{figure}

This comparison clarifies the technical contribution. The controller structure is
classical PSS/residue design; the contribution is applying it to a
DAE-sited data-center load actuator with explicit phase feasibility,
DVFS constraints, and same-system comparison against a local droop
baseline.

\subsection{Multi-mode behavior}
\label{sec:multimode}

\begin{figure}[t]
\centering
\includegraphics[width=\linewidth]{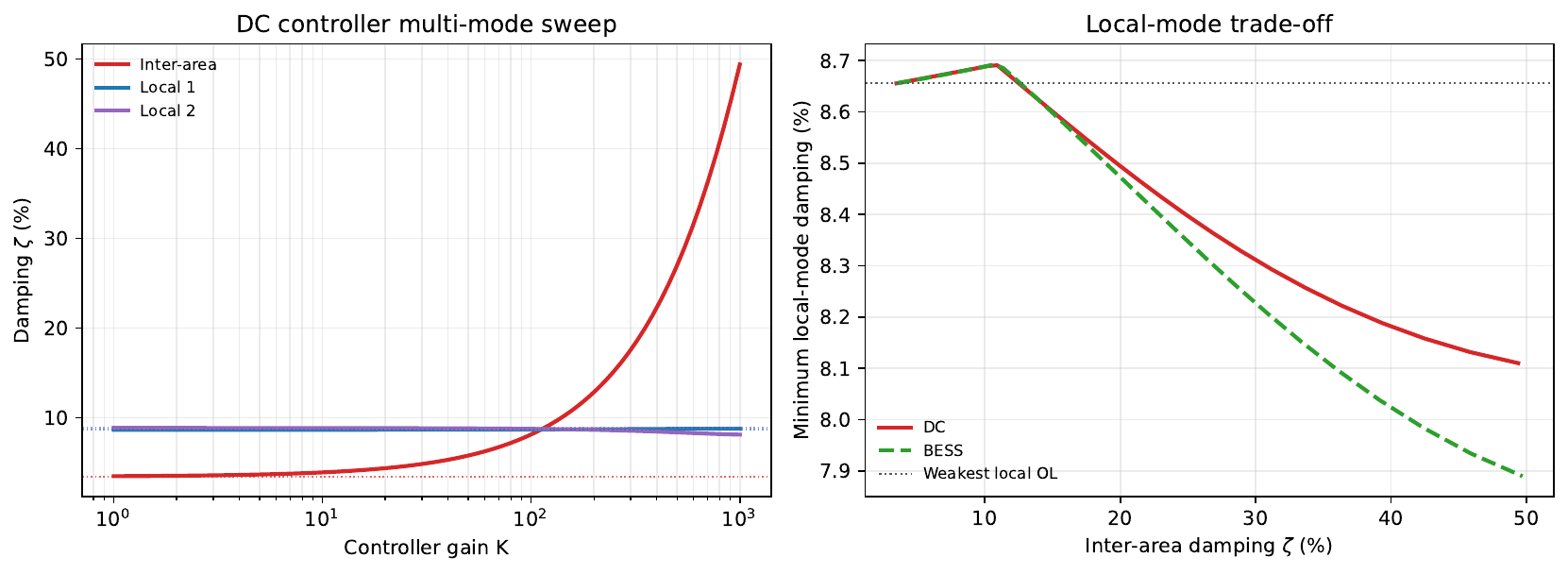}
\caption{Multi-mode behavior on Kundur.
(a) Damping of all three modes vs $K$ (DC).
(b) Trade-off frontier: inter-area damping vs minimum local-mode damping
(DC vs BESS).}
\label{fig:multimode}
\end{figure}

A controller tuned for one mode can de-damp others. Figure~\ref{fig:multimode}
tracks all three Kundur electromechanical modes across the $K$ sweep.

\begin{table}[t]
\caption{Multi-mode damping at selected DC controller gains (Kundur).}
\label{tab:multimode}
\centering
\begin{tabular}{cccc}
\toprule
$K$ & $\zeta_{\mathrm{IA}}$ (\%) & $\zeta_{\mathrm{A1}}$ (\%) & $\zeta_{\mathrm{A2}}$ (\%) \\
\midrule
$0$ (OL)      & $3.43$  & $8.66$ & $8.86$ \\
$9.7$         & $3.89$  & $8.66$ & $8.85$ \\
$51$          & $5.83$  & $8.67$ & $8.80$ \\
$103$         & $8.27$  & $8.68$ & $8.75$ \\
$207$         & $13.18$ & $8.70$ & $8.64$ \\
\bottomrule
\end{tabular}
\end{table}

Inter-area damping rises monotonically with $K$. The weakest local mode
stays within approximately $0.2$ percentage points of the open-loop value
over the practical gain range. Under the DAE-consistent input model, the
Bus~10 WADC does not materially degrade either local mode in the studied
gain range.

\subsection{Synthesis}

The Kundur analyses support a specific conclusion. The DAE-consistent
Bus~10 input has favorable phase, so the 10--50~ms
actuator-lag difference is not enough to create a large peak-MW
advantage over BESS. The robust findings are instead: (i) the DAE
ranking preserves Bus~10 as the best non-generator site, (ii) the
wide-area controller outperforms a local data-center droop analog under
a comparable MW envelope, (iii) the 10~ms actuator has a measurable but
modest latency-margin advantage, and (iv) local modes are not materially
degraded over the practical gain range.

\subsection{Actuator saturation and one-sided DVFS stress tests}
\label{sec:saturation}

DVFS is not naturally symmetric around an uncurtailed baseline: if the
facility is already at full workload, it can reduce load but cannot
increase above nominal. We therefore simulate the controller with hard
MW limits in the reduced DAE closed loop. Four event-equivalent initial
conditions are used: a 1\% modal perturbation, a larger fault-cleared
modal perturbation, a line-outage-equivalent angle split, and a
generator-trip-equivalent speed imbalance. These are reduced-order
stress tests, not electromagnetic-transient fault studies; their purpose
is to test actuator constraints under repeatable DAE-consistent
conditions.

\begin{table}[t]
\caption{Saturation stress-test summary. Decay ratio is the peak
inter-area signal after 15~s divided by the peak over 0.5--5~s.}
\label{tab:saturation}
\centering
\begin{tabular}{lccc}
\toprule
Scenario & Bidirectional peak MW & 2~MW saturation & No-up-headroom decay \\
\midrule
1\% modal & $0.011$ & $0.000$ & $0.059$ \\
Fault-cleared equivalent & $0.068$ & $0.000$ & $0.059$ \\
Line-outage equivalent & $1.505$ & $0.000$ & $0.065$ \\
Generator-trip equivalent & $2.000$ & $0.427$ & $0.098$ \\
\bottomrule
\end{tabular}
\end{table}

\begin{figure}[t]
\centering
\includegraphics[width=\linewidth]{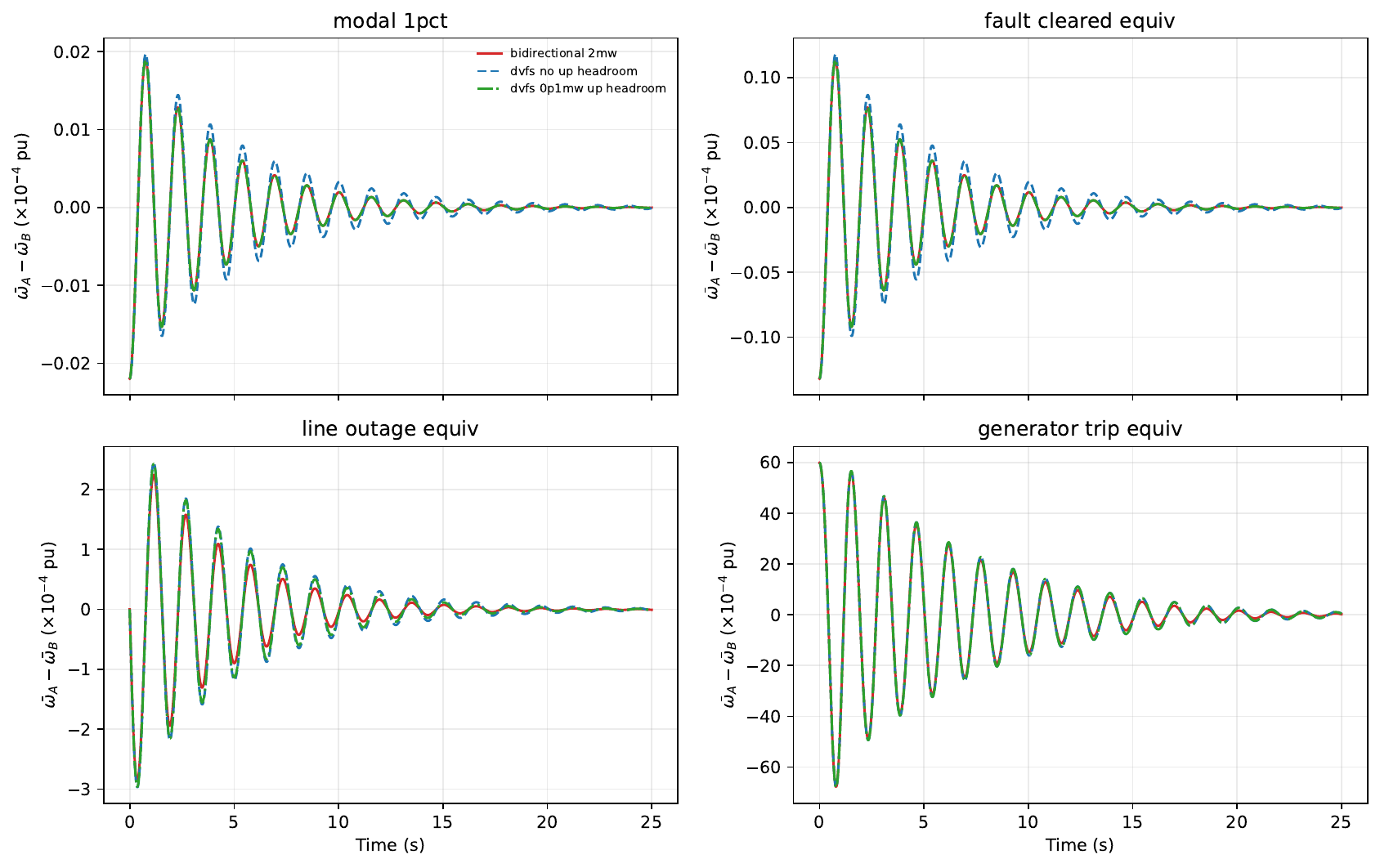}
\caption{Reduced-order saturation and one-sided-DVFS stress tests.
The no-up-headroom case clips positive load commands; a small amount of
precurtailed upward headroom restores the near-bidirectional response
for modal, fault-cleared, and line-outage-equivalent cases.}
\label{fig:saturation}
\end{figure}

The 2~MW envelope is non-binding for modal, fault-cleared, and
line-outage-equivalent cases but binds in the generator-trip-equivalent
case. The no-up-headroom DVFS case approximately doubles the decay ratio
for the smaller disturbances, while $0.1$~MW of upward headroom restores
the bidirectional response for the cases whose positive command remains
below that value. Thus an enrolled data center should operate with a
small precurtailed headroom band if the service requires symmetric
oscillation damping.

\subsection{IEEE 39-bus phase-feasibility scale-up test}
\label{sec:ieee39-scaleup}

We apply the methodology to IEEE 39-bus (10 generators, 46 lines,
230 reduced differential states after conditioning). Because the ANDES-shipped
\texttt{ieee39\_full.xlsx} includes GENROU and IEEEST stabilizers only,
we condition it by inserting EXDC2 exciters and TGOV1 governors for
all 10 generators using textbook parameters~\cite{kundur1994stability},
disable the built-in stabilizers to establish a pre-stabilization
baseline, and rescale $M_i \to M_i/4$ to obtain modes at classical
machine-base frequencies.

The implementation reproduces the ANDES state matrix after zero-time-constant and algebraic-constraint elimination and then applies the same
DAE input construction used for Kundur. The selected multi-generator
mode in this conditioned case is strongly damped
($f=1.423$~Hz, $\zeta=36.23\%$), so IEEE~39 is not used to infer peak-MW
performance. It is used to test the scaling of the bus-ranking and
phase-feasibility procedure.

Algorithm~\ref{alg:siting}'s ranking produces an engineering finding:
the largest non-generator residue occurs at Bus~22
($|R|=4.70\times10^{-3}$), but its residue phase is $-166.1^\circ$,
requiring approximately $+165.4^\circ$ of compensation after the
washout. This exceeds the single-stage lead-lag bound
$\phi_{\max}\approx65^\circ$. In fact, no top-ranked non-generator bus
in the conditioned case satisfies the single-stage phase-feasibility
criterion, indicating that a larger system may require two-stage
compensation or coordinated multi-DC actuation rather than a single
lead-lag block.

\begin{figure}[t]
\centering
\includegraphics[width=\linewidth]{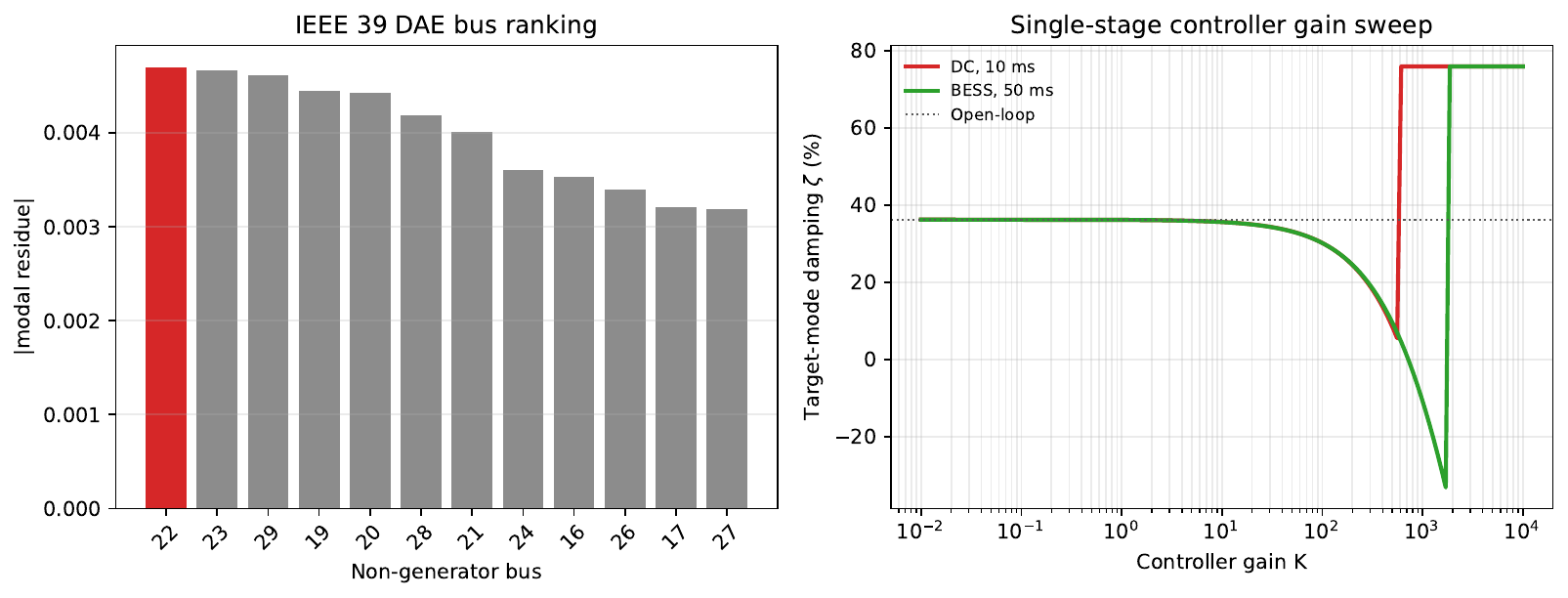}
\caption{IEEE 39-bus scale-up with DAE input. (a) DAE residue ranking
for non-generator buses. (b) Single-stage controller gain sweep showing
the strong dependence on phase feasibility.}
\label{fig:ieee39}
\end{figure}

The overall implication is that the DAE-based ranking procedure generalizes,
but the single-stage controller used on Kundur should not be assumed to
scale automatically. At larger scale, phase feasibility can dominate raw
residue magnitude; practical designs should include either multi-stage
phase compensation or multiple geographically separated data-center
actuators.

\section{Discussion}
\label{sec:discussion}

\subsection{Scope and limitations}

First, the main controller synthesis remains small-signal. The
saturation study tests large initial perturbations and hard MW limits in
the DAE-reduced closed loop, but it is not an electromagnetic-transient
fault study. Second, the absolute peak-MW numbers scale with the chosen
disturbance amplitude; the robust comparison is therefore the
same-disturbance ratio between controllers, not the absolute MW value by
itself. Third, DVFS asymmetry is operationally important. The saturation
simulations show that no-upward-headroom operation clips half-cycles and
that a small precurtailed headroom band is needed for near-bidirectional
damping. Fourth, Kundur has a single dominant inter-area mode; real
interconnections have multiple modes and may require parallel residue
compensation stages or robust-control design. Finally, the IEEE~39 case
used here is a conditioned benchmark with high modal damping; it is
used here to test DAE ranking and phase feasibility rather than to infer
a system-wide performance advantage.

\subsection{Deployment considerations}

\textbf{Communication architecture.} Required PMU feed is available
from existing utility WAMS~\cite{chakrabortty2011wams} at $30$--$80$~ms
end-to-end latency, within DC's stability margin but near BESS's.

\textbf{Workload scheduling.} DVFS modulation of $<1\%$ nameplate over
$20$~s yields latency-factor increases below $1\%$ via
(\ref{eq:dvfs-power}); production schedulers already tolerate this
for thermal management.

\textbf{Thermal time budget.} DVFS can sustain full-range modulation
for $2$--$5$ minutes before cooling dynamics mask IT power changes;
damping transients complete in under $10$~s, well within this budget.

\textbf{Fallback.} If the DC controller loses its feed, the
closed-loop reverts to open-loop; the system is no worse than before
DC deployment. Exponential gain ramp-down over $1$--$2$~s provides
bumpless transfer.

\subsection{Relationship to prior and concurrent work}

End-use-load damping has been studied for over a decade~\cite{liu2014loaddamping,huang2016widearea};
most prior work targets aggregated thermostatic loads with aggregation
bandwidths on the order of seconds. Data-center fast-frequency-response
work, including Al~Kez~\textit{et~al.}~\cite{alkez2021ffr}, establishes
that data centers can provide fast active-power services, but it does
not address inter-area residue siting or phase compensation. Recent
grid-interactive-data-center work~\cite{colangelo2026gridDC} frames AI
data centers as flexible grid assets; this paper studies one specific
oscillation-damping control mechanism within that broader framing.
BESS-based WADC is mature~\cite{kumar2025bess,pal2005robustcontrol};
the comparison here uses BESS as a sensitivity case rather than as a
proxy for a broad peak-MW performance advantage.

\subsection{Future work}

(i) Two-stage lead-lag or $H_\infty$ synthesis for buses with residue
phase beyond single-stage reach; (ii) multi-DC coordinated control for
systems with multiple inter-area modes; (iii) electromagnetic-transient
validation under large-signal faults; (iv) hardware-in-the-loop
demonstration with a DVFS-controlled compute cluster coupled to a
real-time simulator.

\section{Conclusion}
\label{sec:conclusion}

Hyperscale data centers, controlled via DVFS of compute loads, can act
as wide-area damping actuators under a specific set of small-signal and
operating constraints. A
DAE-consistent input matrix preserves the Kundur Bus~10 siting result
and yields a near-damping-only movement of the 0.647~Hz inter-area mode
from 3.43\% to approximately 6\%. The BESS comparison shows comparable
peak-MW requirements for 10--50~ms actuators under the same small-signal
disturbance. The stronger result is the
same-system comparison against a local data-center droop analog: the
wide-area residue controller reaches the 6\% target with about half the
peak modulation. PMU-delay robustness and multi-mode behavior remain
acceptable, and saturation tests identify the need for a small
precurtailed headroom band when DVFS is operated as a symmetric damping
service.

At IEEE~39 scale, the DAE-based ranking procedure exposes a phase-feasibility
constraint that is hidden in the simpler Kundur system: the largest
residue buses can require more phase compensation than a single lead-lag
stage can provide. Future work should therefore focus on multi-DC
coordinated control, two-stage or robust phase compensation,
electromagnetic-transient validation, and hardware-in-the-loop
demonstration with a DVFS-controlled compute cluster.

\section*{Declarations}

\paragraph{Funding.} Not applicable. No external funding was received
for this research.

\paragraph{Conflicts of interest / Competing interests.}
The author declares no financial or non-financial competing interests
related to the content of this manuscript.

\paragraph{Availability of data and material.}
The Kundur 2-area and IEEE 39-bus benchmarks are standard public test
cases distributed with the open-source ANDES power system simulator.
Linearized state matrices, eigenvalue data, participation factors, and
time-domain trajectories supporting the findings of this study are
generated by the scripts described below and reproducible without
additional datasets. Intermediate artifacts are available from the
author on reasonable request.

\paragraph{Code availability.}
Analysis code and figure-generation scripts reproducing all results are
available from the corresponding author on reasonable request, and will
be released under an open-source license upon publication to support
reproducibility.

\paragraph{Authors' contributions.}
M. Shamseldein is the sole author and was solely responsible for
conceptualization, methodology, software, formal analysis,
investigation, visualization, and writing of the original draft,
review, and editing.

\paragraph{Ethics approval.} Not applicable. This research did not
involve human participants or animals.

\paragraph{Consent to participate.} Not applicable.

\paragraph{Consent for publication.} Not applicable.

\bibliography{refs}

\end{document}